
\baselineskip=0.6cm
\magnification=1200
\vskip 8.0truecm
\centerline  {{\bf A CHARACTERIZATION OF THE DIFFERENTIAL}}
\centerline {{\bf IN SEMI-INFINITE COHOMOLOGY}}
\vskip 3.0truecm
\centerline {F\"{u}sun Akman}
\vskip 2.0truecm
\centerline {{\it Department of Mathematics, Yale University,}}
\centerline {{\it New Haven, CT 06520}}
\centerline{{\it akman@pascal.math.yale.edu}}
\vskip 2.0truecm
\centerline {(to appear in {\it Journal of Algebra})}
\filbreak
\centerline {\bf A CHARACTERIZATION OF THE DIFFERENTIAL}
\centerline {\bf IN SEMI-INFINITE COHOMOLOGY}
\vskip 2.0truecm
\centerline {F\"{u}sun Akman}
\centerline {\it Department of Mathematics, Yale University,}
\centerline {\it New Haven, CT 06520}
\vskip 2.0truecm

\centerline{ACKNOWLEDGEMENT}
\vskip 1.0truecm
It is my pleasure to thank Gregg Zuckerman for reading the manuscript
and for the semi-infinitely many comments. His Fall 91 course on
Representations of Infinite Dimensional Lie Algebras at Yale
University restored my faith in mathematical physics, besides providing an
extensive background.
\vskip 2.0truecm
\centerline{INTRODUCTION}
\vskip 1.0truecm
Semi-infinite cohomology is by now firmly established as a means of
describing certain aspects of string theory in modern mathematical
physics [5]. For mathematicians, the first rigorous formulation was
given by Feigin [1], and further clarification and explicit
calculations came with [3]. Several articles ([8],...) have appeared on the
subject (also studied as BRST cohomology), and yet so far, existence
arguments for the differential have been given solely via formulas, or ``by
analogy''. Moreover, one has
to show that the square of the differential is zero each time, the
length of the proof depending on the complexity of the author's
favorite module and on the square of the width of the formula. Hence,
complete proofs practically never get published and semi-infinite
cohomology remains exotic and inaccessible to most mathematicians.

The main goal of this article is to characterize a universal
differential, $d$, which will encode the properties of a given {\bf Z}-graded
Lie algebra $${\cal G}=\oplus _{n \in Z} {\cal G}_{n}$$ with $\dim {\cal
G}_{n} < \infty$ for all n; satisfy the minimal properties that
any self-respecting differential should possess; and induce the
cohomology operator for a very large class of (projective)
representations of ${\cal G}$. The operators of semi-infinite cohomology
theory will be realized as inner derivations of an associative algebra
$Y^{\infty} {\cal G}$, which is (almost) the completed universal
enveloping algebra of the Lie superalgebra $$\tilde{s} ({\cal G}) = \pi
(\tilde{{\cal G}}) \oplus \iota ({\cal G}) \oplus \epsilon ({\cal G}^{\prime})
\oplus Kk$$ where $\tilde{{\cal G}}$ is a certain central
extension of ${\cal G}$, ${\cal G}^{\prime}$ is the restricted dual of
${\cal G}$, and $\iota ({\cal G}) \oplus \epsilon ({\cal G}^{\prime})
\oplus Kk$ ($K$ fixed field of characteristic zero, $k$ central) is the
super Heisenberg Lie algebra based on the underlying vector space of
${\cal G}$. The two gradings $\ast$-deg$=$ superdegree and deg$=$
secondary degree on $\tilde{s} ({\cal G})$, coming from $\ast$-deg
 $\iota (x)=-1$ and $\ast$-deg $\epsilon(x^{\prime})=$ 1 and the grading
on ${\cal G}$ respectively, will induce gradings on $Y^{\infty} {\cal G}$
and its derivations by the Poincar\'{e}-Birkhoff-Witt Theorem (PBW). One
group of such derivations will be $\{ \theta (x)\}_{x \in {\cal G}}$,
corresponding to elements $\{ \underline{ \theta} (x) \}_{x \in {\cal
G}}$ of $Y^{\infty} {\cal G}$, which constitute a natural representation of
${\cal G}$ on $Y^{\infty} {\cal G}$. All these inner derivations will now
act on $Y^{\infty} {\cal G}$ modules of the form $$M \otimes
\wedge^{\infty /2+\ast}{\cal G}^{\prime},$$ {\it M} a smooth
$\tilde{{\cal G}}$ module, $\wedge^{\infty /2+\ast} {\cal
G}^{\prime}$ the module of {\it semi-infinite forms} on ${\cal G}$. Modulo
the precise definitions in the following sections and the choice of a special
homogeneous basis $\{ e_{i} \} _ {i \in Z}$ for ${\cal G}$, the main
result can now be stated as

\vskip 0.8truecm
THEOREM. (a) {\it There exists a unique superderivation} $d$ {\it of}
$Y^{\infty} {\cal G}$ {\it with} $\ast$-{\it deg 1, deg 0, satisfying} $$d
\iota (x)=\underline{\theta} (x) \quad \quad \forall x \in {\cal G}.$$
\vskip 0.5truecm
(b) {\it Moreover},$$d \underline{\theta} (x)=0 \quad \quad \forall x \in {\cal
G}$$ {\it and} $$d^{2} =0.$$
\vskip 0.5truecm
(c) {\it There is a unique element} $\underline{d}$ {\it of the same degrees
in}
$Y^{\infty} {\cal G}$, {\it namely} $$\underline{d} =\sum_{i} \pi (e_{i})
\epsilon ({e_{i}}^\prime)+ \sum_{i<j}:\iota ([e_{i},e_{j}]) \epsilon
({e_{j}}^\prime) \epsilon ({e_{i}}^\prime): \quad ,$$ {\it for which}
$$du=[\underline{d},u],\quad \quad u \in Y^{\infty} {\cal G},$$ {\it and}
$\underline{d}^{2}=0$.
\vskip 0.5truecm
(d) {\it Hence, if} $M$ {\it is any smooth} $\tilde{{\cal G}}$ {\it
module on which the central element acts by 1, then} $\underline{d}$
{\it induces a well-defined differential on the} $Y^{\infty} {\cal G}$
{\it module} $$M \otimes
\wedge^{\infty /2+\ast} {\cal G}^{\prime}.$$ {\it The cohomology}
$H^{\ast}(M \otimes \wedge^{\infty /2}{\cal G}^{\prime} ,\underline{d})$ {\it
is by
definition the semi-infinite cohomology of} ${\cal G}$ {\it with
coefficients in} $M$.
\vskip 1.0truecm
We will show the existence, uniqueness, and nilpotency of a canonical
differential $d$ once and for all, starting with the only piece of
data we need, namely ${\cal G}$, and using only the language of
derivations. In the end, we will obtain the familiar but generic and
-algebraically speaking- well motivated formula that establishes $d$
as an inner derivation. Note that by the Theorem (and by the
definition of completion) $d$ has to yield the classical differential in
case of an ungraded finite dimensional Lie algebra ${\cal G}$.
Hopefully some physics terminology will be demystified in the process.
\vskip 2.0truecm
\centerline {PRELIMINARIES}
\vskip 1.0truecm
Unlike the classical theory, derivations of associative and Lie
algebras have never been utilized systematically in semi-infinite
cohomology. If ${\cal A}$,${\cal B}$ are (super) {\bf Z}-graded associative
or Lie algebras, with a degree-preserving algebra map $\phi : {\cal A}
\rightarrow {\cal B}$, a {\it superderivation} $D: {\cal A}
\rightarrow {\cal B}$ of $\ast$-deg N with respect to $\phi$ is
a linear map of $\ast$-deg N satisfying $$D(u \cdot v)=(Du)\cdot (\phi
v)+(-1)^{N(\ast deg u)} (\phi u) \cdot (Dv)$$ for $u,v \in {\cal A}$,
$u$ homogeneous. In case of secondary {\bf Z}-gradings also preserved by
$\phi$, we will still talk about a superderivation of $\ast$-deg M and
(secondary) deg N even though this new grading may not be compatible
with the ``super'' structure. Note that any associative algebra
derivation is a Lie algebra derivation in a natural way. The converse
is not true.

A reliable way of constructing a derivation is extending another
derivation, or just a linear map, to an associative algebra generated
in some way by the original space. For example, any linear map $f:E
\rightarrow {\cal A}$ into an associative algebra extends to an
algebra derivation $f:TE \rightarrow {\cal A}$ from the tensor algebra,
if we first define p-linear maps $$E \times \cdots \times E
\rightarrow {\cal A}$$ with $$(u_{1},\ldots ,u_{p}) \mapsto
\sum_{k=1}^{p} \phi (u_{1}) \cdots f(u_{k}) \cdots \phi (u_{p}).$$
Here $\phi :E \rightarrow {\cal A}$ is a given linear map and $\phi
:TE \rightarrow {\cal A}$ the corresponding algebra map. Using this
prototype as an intermediate step, a Lie algebra derivation $f:\ell
\rightarrow {\cal A}$ into an associative algebra can be extended to
an associative algebra derivation $f: {\cal U}\ell \rightarrow {\cal A}$
from the universal enveloping algebra (again $\phi :\ell \rightarrow
{\cal A}$ given Lie algebra map, with extension $\phi :{\cal U} \ell
\rightarrow {\cal A}$ by the universal property of ${\cal U} \ell$) :
The derivation $f:T \ell \rightarrow {\cal A}$ factors through $T \ell
\rightarrow {\cal U} \ell =T \ell / ideal \quad$ since elements of the form
$$u \otimes (x \otimes y - y \otimes x - [x,y])\otimes v =u \otimes
([x,y]_{T \ell} -[x,y]_{\ell})\otimes v$$ with $x,y \in \ell, u,v \in
T \ell$ are annihilated. The result also holds for super Lie algebras,
so we immediately get extensions of linear maps $E \rightarrow {\cal
A}$ to associative algebra derivations $SE \rightarrow {\cal A}$ and
$\wedge E \rightarrow {\cal A}$ (here $SE,\wedge E$ are the symmetric and
exterior algebras) as special cases. Once we define completions of
certain enveloping algebras, our Lie derivations will extend uniquely
to the completion and we will not distinguish between the generating map and
the full one by virtue of this discussion. Derivations form a
super Lie algebra themselves under the obvious superbracket. The map
$\phi$ will be an inclusion most of the time, and we will sometimes
omit the prefix {\it super}-.

Let us now define the completion of a {\it tame} {\bf Z}-graded (super) Lie
algebra $\ell$ with a secondary {\bf Z}-grading $$\ell = \oplus_{n}
\ell_{n}.$$ ``Tame'' means each $\ell_{n}$ is finite dimensional (the
terminology is Zuckerman's). Let $$\ell_{\leq 0}=\oplus_{n \leq 0}
\ell_{n}$$ and $$\ell_{+}=\oplus_{n>0} \ell_{n}.$$ By PBW their
universal enveloping algebras are related by $${\cal U}\ell \cong
{\cal U} \ell_{\leq 0} \otimes_{K} {\cal U} {\ell}_{+}$$ as vector spaces.
The grading in ${\cal U} {\ell}_{+}$ induces a subsidiary grading which we
denote by sdeg, as opposed to deg, which comes from the total
(secondary) grading on $\ell$. A typical homogeneous element of ${\cal
U}\ell$ is of the form $$\sum_{k=1}^{r} u_{k} v_{k}$$ with $u_{k}
\in {\cal U}\ell_{\leq 0}, v_{k} \in {\cal U} \ell_{+}$, and
deg$(u_{k} v_{k})=$deg$(u_{l} v_{l})$ $\forall$ $k,l$. If infinite
expressions $$\sum_{k=1}^{\infty} u_{k} v_{k}$$ with
deg$v_{k}=$sdeg$(u_{k} v_{k}) \rightarrow \infty$ as $k \rightarrow
\infty$ are also allowed, one gets a canonical completion ${\cal
U}^{\infty} \ell$ of ${\cal U} \ell$ with respect to deg [10]. Completions
of universal enveloping algebras have appeared in the work of Kac [6]
and Zhelobenko [9], for example, but the construction is somewhat different
 and the grading used is the so-called primary grading for
Kac-Moody algebras. Probably the object closest to this one in
literature is the ``universal enveloping algebra of a vertex operator algebra''
in the Frenkel-Zhu paper [4], where infinite sums are
built out of modes of vertex operators.
 In general, it can be shown that
products exist and are exactly what we think they ought to be [10].
${\cal U}^{\infty} \ell$ is a topological associative algebra with
${\cal U}\ell$ as a dense subalgebra, and both the multiplication and
the induced Lie bracket are continuous. Note that $${\cal U}^{\infty}
\ell =\oplus_{n} ({\cal U} \ell_n)^{\infty}$$ by definition, that is,
each homogeneous piece is completed separately.

${\cal U}^{\infty}\ell$ provides a context for the puzzling infinite
sums of mathematical physics in terms of elements of $\ell$. The
normal ordering : : that appears in most sums merely (another
algebraist's belittlement) ensures the validity of such expressions as
elements of the completed algebra. What we really want to do with them
is to take their supercommutators, and : : is superfluous once inside
the bracket. Since ${\cal U}^{\infty} \ell$ acts on every {\it smooth}
$\ell$ module (a {\bf Z}-graded $\ell$ module where each ${\cal U} {\ell}_{+}$
orbit is finite dimensional) by its very definition, Lie algebra maps
$${\cal G} \rightarrow {\cal U}^{\infty} {\cal H},$$ which abound in
physics (${\cal G}=$ constraint algebra, ${\cal H}=$ current algebra),
are used to generate an action of ${\cal G}$ on smooth ${\cal H}$
modules. Such a map always extends to $${\cal U}^{\infty} {\cal G}
\rightarrow {\cal U}^{\infty} {\cal H}$$ by abstract nonsense. More
about this later.

At this point we reluctantly fix a homogeneous basis $\{ e_{i} \}_{i
\in Z}$ for ${\cal G}$ such that whenever $e_{i} \in {\cal G}_{n}$,
either $e_{i+1} \in {\cal G}_{n}$ or $e_{i+1} \in {\cal G}_{n+1}$, with
the ultimate purpose of generating formulas. What precedes that will
be basis independent. Following the notation of [1], [3], and [10],
let $${\cal G}^{\prime} = \oplus_{n} {\cal G}_{n}^{\prime}$$ be the
{\it restricted dual} of ${\cal G}$, with $${\cal G}_{n}^{\prime} =
Hom_{K}({\cal G}_{-n},K).$$
Then $\{ e_{i}\prime \}$ is the dual basis in ${\cal G}\prime$. Let
$$s=s({\cal G})={\cal G}\oplus {\cal G}\prime \oplus Kk $$
be the super Heisenberg Lie algebra with center $Kk$ in which elements
of ${\cal G}$ and ${\cal G}\prime$ will be distinguished by the
additional symbols $\iota$ and $\epsilon$, emphasizing that ${\cal G}$
and ${\cal G}\prime$ are now the adjoint and coadjoint
representations. The $\ast$-degrees for elements of ${\cal G}$, ${\cal
G}\prime$, and $Kk$ are $-1$, $1$, and $0$ respectively, and both
commutators and anticommutators will be denoted by the superbracket
$[\quad , \quad]$. Thus
$$[\iota(x),\iota(y)]=[\epsilon(x\prime),\epsilon(y\prime)]=0 $$
and
$$[\epsilon(x\prime), \iota(y)]=<x\prime , y>k $$
for all $x,y \in {\cal G}$, $x\prime , x\prime \in {\cal G}\prime$. A
secondary {\bf Z}-grading is given by
$$ \deg \iota(x)=n, \quad x \in {\cal G}_{n}, \quad \deg (x
\prime)=-n, \quad x\prime \in {\cal G}_{n}\prime \quad \deg k=0. $$
We will use the same letter to denote dual elements, e.g. $\iota(x)$
and $\epsilon(x\prime)$.

In the abstract bigraded algebra $s$, the two compatible gradings have
different
uses: The super identities $$[u,v]+(-1)^{(\ast \deg u)(\ast \deg
v)}[v,u]=0$$ and $$(-1)^{(\ast \deg u)(\ast
\deg w)}[u,[v,w]]$$  $$ +(-1)^{(\ast \deg v)(\ast \deg
u)}[v,[w,u]]$$  $$+(-1)^{(\ast\deg w)(\ast \deg v)} [w,[u,v]]=0 $$
hold for $\ast$-deg, and we complete ${\cal U}s$ with respect to deg.
Let $$C^{\infty} {\cal
G}={\cal U}^{\infty} s /(k=1),$$ the completed Clifford algebra on
${\cal G}$. Then by PBW $$C^{\infty} {\cal G} \cong (\wedge {\cal G}
\otimes \wedge {\cal G}^{\prime})^{\infty}$$ as vector spaces.
(Another approach would be to define $C^{\infty} {\cal G}$ to be
${\cal U}^{\infty} s [1/k]$ and modify all formulas by a factor of $1/k$.)

$\wedge^{\infty /2 +\ast} {\cal G}^{\prime}$ is the $C {\cal G}={\cal
U} s ({\cal G}) / (k=1)$ module induced from the one dimensional trivial
representation of the subalgebra $$\wedge ({\cal G}_{\geq 0} \oplus
{\cal G}_{-}^{\prime}),$$ whose elements are called {\it annihilation
operators}. This module was first realized as the span of the {\it
semi-infinite forms} $$e_{i_{1}}^{\prime} \wedge e_{i_{2}}^{\prime}
\wedge \cdots \wedge e_{i_{k}}^{\prime} \cdots$$ where $i_{1} > i_{2}
> \cdots $ and $i_{k+1} = i_{k} -1$ from some point on. The completed
algebra also acts on $\wedge^{\infty /2 +\ast}{\cal G}^{\prime}$.

The following results, together with Lemma 3, an extension of the
second one, essentially form the proof of the Theorem. As an
unexpected bonus, they help one guess and verify formulas for
derivations, as well as take commutators of normal ordered expressions
easily.
\vskip 0.8truecm
LEMMA 1. {\it The center of} $C^{\infty} {\cal G}$ {\it is} $K$.

{\it Proof.} Any PBW monomial in $C^{\infty} {\cal G}$ is uniquely
determined by its commutators with all the $\iota (x)$ and $\epsilon
(y^{\prime})$: If there exists an $\epsilon (e_{i}^{\prime})$ in the
monomial, $[\iota (e_{i}), \quad ]$ replaces it with $\pm$1. Otherwise
it kills the monomial. Similarly $[\epsilon (e_{i}^{\prime}), \quad ]$
replaces an existing $\iota (e_{i})$ by $\pm$1 or else kills the
monomial. A central element commutes with all $\iota (x)$ and
$\epsilon (y^{\prime})$, hence cannot contain a nonconstant monomial
as a summand. So {\it Center} $C^{\infty} {\cal G}=K$, of $\ast$-deg and deg
0.
\vskip 0.8truecm
LEMMA 2. {\it Let} $D: s \rightarrow C^{\infty} {\cal G}$
({\it equivalently}, $C^{\infty} {\cal G} \rightarrow C^{\infty} {\cal G}$)
{\it be a superderivation of} $\ast$-{\it degree N}$\geq 0$. {\it Then} $D$
{\it is
determined by its values on} $\iota ({\cal G}) \oplus Kk \subset
s({\cal G})$.

{\it Proof.} A linear map $D:s \rightarrow C^{\infty} {\cal G}$ of
$\ast$-degree N is a derivation iff $$D[u,v]_{s}=[Du,v]_{C^{\infty}
{\cal G} }+ (-1)^{N(\ast \deg u)}[u,Dv]_{C^{\infty} {\cal G}}$$ for
all $u,v \in s$, $u$ homogeneous. If $Dk=0$, this is reduced to
$$[Du,v]=(-1)^{N(\ast \deg u)+1} [u,Dv].$$ Let $D_{1},D_{2}$ be
derivations of $\ast$-deg N such that $D_{1}=D_{2}$ on ${\cal G}
\oplus K$. Then for $D=D_{1} -D_{2}$, we have $$[D \epsilon
(x^{\prime}), \iota (y)]= (-1)^{-N+1} [\epsilon (x^{\prime}), D \iota
(y)]=0 \quad \forall x^{\prime} \in {\cal G}^{\prime}, y \in {\cal
G}.$$ This shows that $D \epsilon (x^{\prime})$, which is of
$\ast$-deg N$+1 \geq 0$, does not have any $\epsilon$'s. Constants are
also excluded by degree considerations so $D \epsilon
(x^{\prime})=0$. Therefore $D_{1}=D_{2}$ on $s$.
\vskip 0.8truecm
{\it Remarks.} (1) A similar result holds for $\ast$-degree N$\leq 0$
and ${\cal G}^{\prime} \oplus Kk$.

(2) Since $C^{\infty} {\cal G}$ is generated by $s$, the extension of
any derivation of $s$ into $C^{\infty} {\cal G}$ carries the center to
the center. Hence any derivation of nonzero $\ast$-degree is trivial
on $Kk$.
\vskip 0.8truecm
The new techniques can best be demonstrated by constructing an action
of ${\cal G}$ (rather, of a central extension) on $C^{\infty} {\cal
G}$ by inner derivations. The action $$ \iota (y) \mapsto \iota (ad
(x) \cdot y), \quad \epsilon (y^{\prime}) \mapsto \epsilon (
ad^{\prime} (x) \cdot y^{\prime}), \quad 1 \mapsto 0$$ of $x \in
{\cal G}$ in $s$ extends to a derivation of $C^{\infty} {\cal G}$ by
previous arguments. If we can produce an element, say $\underline{\rho}
(x)$, such that $\rho (x)=[\underline{\rho} (x), \quad]\quad $ has the
correct degrees and agrees with the action of $x$ on $\iota ({\cal G})
\oplus Kk$, then the extension is exactly $\rho (x)$ by Lemma 2. (We
will consistently underline elements corresponding to inner
derivations.) Our candidate is $$\underline{\rho} (x) = \sum_{i \in Z} :
\iota (ad ( x) \cdot e_{i}) \epsilon (e_{i}^{\prime}) : \quad \in
C^{\infty} {\cal G}$$ (the formula looks slightly different than
the traditional one, say in [3], but this order is more natural and
leads directly to the correct formula for $\underline{d}$).
The normal ordering , : :, means a factor with
positive degree should appear on the right, and if we have to change
the given order, we had better multiply that term by $-1$. If we take
the commutator of $\underline{\rho} (x)$ with an element of $s$, we
will be allowed to do it term by term by continuity, and the central
term arising from the possible change of order will vanish. The sum
over {\bf Z} is shorthand for two different sums over nonnegative
integers. Also recall that $$\deg \epsilon (e_{i}^{\prime})=
\deg e_{i}^{\prime}= - \deg e_{i}$$ and $$\deg \iota
(ad (x) \cdot e_{i})=\deg [x,e_{i}]=\deg x + \deg
e_{i},$$ so that $\underline{\rho} (x)$ is a valid expression.
{}From now on we will write such sums with a clear conscience and not
offer explanations. We immediately check that $$[\underline{\rho}
(x),\iota (y)]=\iota (ad (x) \cdot y) \quad \quad x,y \in {\cal G},$$
hence $\rho (x)$ and the action of $x$ are the same derivation by
Lemma 2. Here is part of the computation: $$\eqalign{ [:\iota (ad
(x) \cdot e_{i}) \epsilon (e_{i}^{\prime}):,\iota (e_{j})] &=
[\iota (ad (x) \cdot e_{i}) \epsilon (e_{i}^{\prime}), \iota
(e_{j})] \cr {}&= \iota (ad (x) \cdot e_{i}) [\epsilon
(e_{i}^{\prime}), \iota (e_{j})] \cr {}&= \iota (ad (x) \cdot
e_{i}) \delta_{ij}. }$$ Of course, addition of central terms to
$\underline {\rho} (x)$ will not change $\rho (x)$, but we always
impose the condition that deg $\underline{\rho} (x)$ be the same as
that of $x$. By Lemma 1, $\rho (x)$ is uniquely represented by
$\underline{\rho} (x)$ as long as deg $x \neq 0$.

The span $\{ \rho (x) \} \subset DerC^{\infty} {\cal G}$ closes as a Lie
algebra, but $\{ \underline{\rho} (x) \} \subset C^{\infty} {\cal G}$
does not in general. We have $$[[\underline{\rho} (x), \underline{\rho}
(y)],\iota (z)]=[\underline{\rho} ([x,y]),\iota (z)]=\iota
([[x,y],z])$$ for all $x,y,z \in {\cal G}$. For reasons to become
clear later we would like to have $$\gamma (x,y) =_{def}
[\underline{\rho} (x),\underline{\rho} (y)]-\underline{\rho}
([x,y])=0$$ for $x,y \in {\cal G}$. It is easy to see that $\gamma$
is a cocycle in the classical sense, i.e $$\gamma (x,[y,z])+\gamma
(y,[z,x])+\gamma (z,[x,y])=0.$$ The calculation makes use of the
Jacobi identity as well as the $\ast$-deg of $\underline{\rho} (x)$
(zero). We can manage to have $\gamma \equiv 0$ when $H^{2} {\cal
G}=0$ (e.g. ${\cal G}=Virasoro$, or ${\cal G}=\ell \otimes K[t,t^{-1}]
\oplus Kc$, affine Kac-Moody algebra) if we redefine
$\underline{\rho}$ by $$\underline{\rho} (x)=\sum_{i} :\iota (ad
(x) \cdot e_{i}) \epsilon (e_{i}^{\prime}):+\beta (x),$$ where $\beta$
is a special element of ${\cal G}_{0}^{\prime}$ ([3], [7] Chapter
7). On the other hand, one can choose to leave $\underline{\rho} (x)$
as it is, for example when ${\cal G}=Witt=DerK[t,t^{-1}]$ or ${\cal
G}= \ell \otimes K[t,t^{-1}]$ might be more appropriate than their
famous central extensions. We will adopt the second method and make up
for the inconvenience later.

On a final note, one can always pretend that the formula for
$\underline{\rho} (x)$ is inspired by Lemma 2, where $\epsilon
(e_{i}^{\prime})$ deletes $\iota (e_{i})$, to be replaced by $\iota
([x,e_{i}])$.
\vskip 2.0truecm
\centerline {FIRST STEP}
\vskip 1.0truecm
We will start by characterizing the part of $d$ that acts on
$C^{\infty} {\cal G}$, namely $d_{0}$. It will be the unique
derivation of $C^{\infty} {\cal G}$ of $\ast$-deg 1, deg 0 mapping
$\iota (x)$ to $\underline{\rho} (x)$. It suffices to define
$d_{0}$ on $s({\cal G})$, i.e. we have to figure out the image of
$\epsilon (x^{\prime})$. This differential only has to satisfy
$$\eqalign{ [d_{0}\iota (x),\iota (y)] &= [\iota (x),d_{0} \iota (y)]
\quad (1) \cr [d_{0}\epsilon (x^{\prime}),\epsilon (y^{\prime})] &=
[\epsilon (x^{\prime}),d_{0}\epsilon (y^{\prime})] \quad (2) \cr
[d_{0} \epsilon (x^{\prime}),\iota (y)] &= [\epsilon (x^{\prime}),d_{0}
\iota (y)] \quad (3)}$$ for all $x,y \in {\cal G},
x^{\prime},y^{\prime} \in {\cal G}^{\prime}$. Condition (1) is
satisfied since $d_{0}\iota (x)=\underline{\rho} (x)$. Next, we
assume $$[d_{0} \epsilon (x^{\prime}), \epsilon (y^{\prime})]=0 \quad
\forall x^{\prime},y^{\prime} \in {\cal G}^{\prime} \quad (4)$$ and
hope that this works. After all, $d_{0} \epsilon (x^{\prime})$ is of
$\ast$-deg 2 and every term has two more $\epsilon$'s than $\iota$'s.
(4) merely says that there are exactly two $\epsilon$'s and no
$\iota$'s. So $$d_{0}\epsilon (x^{\prime})=\sum_{\deg e_{i}+\deg
e_{j}=\deg x} \lambda_{ij} :\epsilon (e_{i}^{\prime}) \epsilon
(e_{j}^{\prime}):\quad , \quad \lambda_{ij} \in K,$$ which can be completely
determined if all $[d_{0}\epsilon (x^{\prime}),\iota (y)]$ are known.
But then we simply define $$\eqalign{[d_{0} \epsilon
(x^{\prime}),\iota (y)] &= [\epsilon (x^{\prime}), d_{0}\iota (y)] \cr
{}&= [\epsilon (x^{\prime}),\underline{\rho} (y)] \cr {}&= -\epsilon
(ad^{\prime} (y) \cdot x^{\prime})}$$ via condition (3). Incidentally, this
gives us
the formula $$d_{0} \epsilon
(x^{\prime}) = -1/2 \sum_{i}:\epsilon (ad^{\prime} (e_{i}) \cdot
x^{\prime}) \epsilon (e_{i}^{\prime}):\quad .$$
The factor $1/2$ is there because there are two $\epsilon$'s for an $\iota$ to
cross. We have $$-1/2 \quad [\sum_{i}:\epsilon (ad^{\prime}
(e_{i}) \cdot x^{\prime}) \epsilon (e_{i}^{\prime}):,\iota
(e_{j})]=-\epsilon (ad^{\prime} (e_{j}) \cdot x^{\prime})$$  as
the result of some mild linear algebra, which verifies our guess.

Now $d_{0}^{2}=1/2[d_{0},d_{0}]$ is a superderivation of $C^{\infty}
{\cal G}$ of $\ast$-deg 2, deg 0, and is determined by its values on
${\cal G}$. Then $$d_{0}^{2}=0 \quad \Leftrightarrow \quad d_{0}^{2} \iota
(x)=0
\quad \forall x \in {\cal G} \quad \Leftrightarrow \quad d_{0} \underline{\rho}
(x) =0 \quad \forall x \in {\cal G},$$ but $d_{0} \underline{\rho}
(x)$ is in turn an element of $\ast$-deg 1 and again Lemma 2 applies:
$$\eqalign{ [d_{0}\underline{\rho} (x) , \iota (y)] &=
d_{0}[\underline{\rho} (x),\iota (y)]-[\underline{\rho} (x),d_{0}\iota
(y)] \cr {}&= d_{0} \iota ([x,y])-[\underline{\rho} (x),
\underline{\rho} (y)] \cr {}&= \underline{\rho}
([x,y])-[\underline{\rho} (x), \underline{\rho} (y)] \cr {}&= - \gamma
(x,y).}$$ So $d_{0}^{2}=0$ is equivalent to the closure of $\{
\underline{\rho} (x) \}_{x \in {\cal G}}$, which was remarked upon
earlier. This approach makes it unnecessary to square giant sums.

A formula for $d_{0}$? Why not. We find an inner derivation of
$C^{\infty} {\cal G}$ of $\ast$-deg 1 and  deg 0, which is the same as
$d_{0}$ on ${\cal G}$. Let $$ \eqalign { \underline{d_{0}}  &= 1/2
\sum_{i} :\underline{\rho} (e_{i}) \epsilon (e_{i}^{\prime}): \cr {}&= 1/2
\sum_{i \neq j} : \iota ([e_{i},e_{j}]) \epsilon (e_{j}^{\prime})
\epsilon (e_{i}^{\prime}): \cr {}&= \sum_{i<j} : \iota ([e_{i},e_{j}])
\epsilon (e_{j}^{\prime}) \epsilon (e_{i}^{\prime}):}$$ (add $\epsilon
(\beta)$ if $\underline{\rho} (x)$ is modified by $\beta (x)$ for
deg$x=0$ ).

Then $$\eqalign {[ \underline{ d_{0} }, \iota (e_{r})] &= 1/2
\sum_{i \neq j} : [ \iota ([ e_{i},e_{j}]) \epsilon (e_{j}^{\prime})
\epsilon (e_{i}^{\prime}), \iota (e_{r}) ] : \cr {}&= 1/2 \sum_{i
\neq j} : \iota ([e_{i},e_{j}]) \{ \delta_{ir} \epsilon
(e_{j}^{\prime}) - \delta_{jr} \epsilon (e_{i}^{\prime}) \} : \cr
{}&= 1/2 \{ \sum_{j} : \iota ([e_{r},e_{j}]) \epsilon (e_{j}^{\prime}) :
- \sum_{i} : \iota ([e_{i},e_{r}]) \epsilon (e_{i}^{\prime}) : \} \cr
{}&= 1/2 \{ \underline{\rho} (e_{r}) + \underline{\rho} (e_{r}) \}
= \underline{\rho} (e_{r}).}$$
\vskip 2.0truecm
\centerline{FINAL TOUCHES}
\vskip 1.0truecm
The cocycle $- \gamma (x,y)$ determines a central extension of ${\cal
G}$, say $\tilde{{\cal G}}$. We extend the chosen basis of ${\cal G}$
to $\{ e_{i} \} \cup \{ c \}$, but now $$[e_{i},e_{j}]_{\tilde{{\cal
G}}} = [e_{i},e_{j}]_{{\cal G}}- \gamma (e_{i},e_{j})c.$$ Form the Lie
superalgebra $\tilde{s}$ as the {\it direct} sum $$\tilde{s} =
\tilde{s} ({\cal G})= \tilde{{\cal G}} \oplus s ({\cal G}) $$ where
$\tilde{{\cal G}}$ is of $\ast$-deg 0 and the rest is as before.
Elements of $\tilde{{\cal G}}$ will be denoted by $\pi (x)$ and will
have secondary grading induced from ${\cal G}$, with deg$\pi (c)=0.
\quad \tilde{{\cal G}}$ will be ``itself'' in $\tilde{s}$ unlike
${\cal G} \subset s({\cal G})$, that is , $$ [\pi (x), \pi
(y)]_{\tilde{s}}= \pi ([x,y]_{\tilde{\cal G}}).$$ Define $$
Y^{\infty} {\cal G} = {\cal U}^{\infty} \tilde{s} ({\cal G})/ (k=1,
\pi (c)=1) $$ which is isomorphic to $$ ({\cal U} \tilde{{\cal G}}
\otimes \wedge {\cal G} \otimes \wedge {\cal
G}^{\prime})^{\infty}$$ by PBW. We will consider derivations $D: \tilde{s}
\rightarrow Y^{\infty} {\cal G}$ and imitate the former construction.
\vskip 0.8truecm
LEMMA 3. {\it Let} $D: \tilde{s} \rightarrow Y^{\infty} {\cal G}$ {\it be a
superderivation of} $\ast$-{\it deg N}$\geq 1$ ({\it in particular} $k \mapsto
0$).
{\it Then} $D$ {\it is determined by its values on} ${\cal G}$ {\it only}.

{\it Proof.} Let $D_{1}=D_{2}$ on ${\cal G}$, and $D=D_{1}-D_{2}$. We
want to show $$D \epsilon (x^{\prime})=0, \quad D \pi (x)=0 \quad
\quad \forall x \in {\cal G}, x^{\prime} \in {\cal G}^{\prime}.$$
Again, $$[D \epsilon (x^{\prime}), \iota (y)]= \pm [\epsilon
(x^{\prime}), D \iota (y)]=0.$$ But $D \epsilon (x^{\prime})$ has at
least two $\epsilon$'s in each term, so $D \epsilon (x^{\prime})=0$.
Similarly, from $[\pi (x), \iota (y)]=0$, we get $$ [D \pi (x), \iota
(y)]= - [\pi (x), D \iota (y)]=0,$$ and $D \pi (x)$, with $\ast$-deg at
least 1, has no $\epsilon$'s. Then $D \pi (x)=0$ and $D_{1}=D_{2}$.
\vskip 0.8truecm
Define $$\underline{\theta} (x)= \pi (x)+ \underline{\rho} (x) \quad \quad
\forall
x \in {\cal G},$$ where $\underline{\rho} (x)$ has no $\beta (x)$
term. This time $$ \eqalign{ [\underline{\theta} (x),\underline{\theta}
(y)] &= [\pi (x), \pi (y)]+[\underline{\rho} (x),\underline{\rho}
(y)] \cr {}&= \pi ([x,y]_{\tilde{{\cal G}}})+\underline{\rho}
([x,y])+ \gamma (x,y) \cr {}&= \pi ([x,y]_{\cal G})-\gamma (x,y) \pi
(c) + \underline{\rho} ([x,y]) + \gamma (x,y) \cr {}&=
\underline{\theta} ([x,y])+\gamma (x,y)(1- \pi (c)) \cr {}&=
\underline{\theta }([x,y]), } $$ which is the phenomenon known as {\it
cancellation of anomalies}. The corresponding inner derivations induce a
genuine action of the graded Lie algebra ${\cal G}$ on modules $M
\otimes \wedge^{\infty /2 + \ast } {\cal G}^{\prime}$, and this
will give us a square-zero differential.
\vskip 2.0truecm
\centerline{PROOF OF THE THEOREM}
\vskip 1.0truecm
As before, we define $d$ on $\epsilon (x^{\prime})$ and $\pi (x)$
only, making sure it is a derivation. Again assume $\quad [d\epsilon
(x^{\prime}),\epsilon (y^{\prime})]=0 \quad $, and define $d \epsilon
(x^{\prime})$ via $$ [d \epsilon (x^{\prime}), \iota (y)]=[\epsilon
(x^{\prime}) , d \iota (y)]= - \epsilon (ad^{\prime}(y) \cdot
x^{\prime}),$$ and $d \pi (x)$ via $$ \eqalign{ [d \pi (x), \iota (y)]
&= -[\pi (x),d \iota (y)] = -[\pi (x), \underline{\theta}(y)] \cr {}&=
-[\pi (x),\pi (y)]= - \pi ([x,y]_{\tilde{\cal G}}), } $$ thanks to
Lemma 3. Then $d \epsilon (x^{\prime})$ is the same as $d_{0} \epsilon
(x^{\prime})$ and $$d \pi (x)= - \sum_{i} \pi ([x,e_{i}]_{\tilde{\cal
G}}) \epsilon (e_{i}^{\prime}).$$ It remains to check the relations $$
[d \pi (x), \epsilon (y^{\prime})]+[\pi (x),d \epsilon (y^{\prime})]=0
\quad \quad (7)$$ and $$\eqalign{ [d \pi (x), \pi (y)]+[ \pi (x), d
\pi (y)] &= d[\pi (x), \pi (y)] \cr {}&= d \pi ([x,y]_{\tilde{\cal
G}}) \quad \quad (8). } $$ Eq.(7) is just $0+0=0$. As for (8),
$$- \sum_{i} \pi ([[x,e_{i}]_{\tilde{\cal G}} ,y]_{\tilde{\cal G}})
\epsilon (e_{i}^{\prime}) -\sum_{i} \pi
([x,[y,e_{i}]_{\tilde{\cal G}}]_{\tilde{\cal G}}) \epsilon
(e_{i}^{\prime}) $$ $$= - \sum_{i} \pi ([[x,e_{i}]_{\tilde{\cal
G}},y]_{\tilde{\cal G}} + [[e_{i},y]_{\tilde{\cal G}},x]_{\tilde{\cal
G}} )  \epsilon (e_{i}^{\prime})$$ $$=- \sum_{i} \pi
([[x,y]_{\tilde{\cal G}} , e_{i}]_{\tilde{\cal G}} ) \epsilon
(e_{i}^{\prime})$$ by the
Jacobi identity.

Once more, $$d^{2}=0 \quad \quad and \quad \quad d \underline{\theta} (x)=0$$
follows from: $$\eqalign{ [d \underline{\theta} (x),\iota (y)] &=
d[\underline{\theta} (x), \iota (y)]-[ \underline{\theta} (x),d \iota
(y)] \cr {}&= d[\underline{\rho} (x), \iota (y)]-[ \underline{\theta}
(x), \underline{\theta} (y)] \cr {}&= d \iota ([x,y])-[
\underline{\theta} (x), \underline{\theta} (y)] \cr {}&=
\underline{\theta} ([x,y])-[ \underline{\theta} (x),
\underline{\theta} (y)] \cr {}&= 0  .} $$ Finally, $$\eqalign{
[\underline{d}, \iota (e_{r})] &= \pi (e_{r})+[\underline{d_{0}} ,
\iota (e_{r})] \cr {}&= \pi (e_{r}) + \underline{\rho} (e_{r}) \cr
{}&= \underline{\theta} (e_{r}), } $$ which shows $d$ is an inner derivation
on $Y^{\infty} {\cal G}$; moreover $\underline{d}$ is unique subject
to the degree conditions. Since
$\underline{[D_{1},D_{2}]}=[\underline{D_{1}}, \underline{D_{2}}]$ in
general, we have $\underline{d}^{2}=0$.
\vskip 2.0truecm
\centerline{AN EXAMPLE: THE SEMI-INFINITE WEIL COMPLEX}
\vskip 1.0truecm
The {\it semi-infinite Weil complex} $W^{\infty /2} {\cal G}$
associated to a {\bf Z}-graded tame Lie algebra ${\cal G}$ is the tensor
product of the semi-infinite symmetric and exterior modules $S^{\infty
/2} {\cal G}^{\prime}$ and $\wedge^{\infty /2} {\cal G}^{\prime}$
together with the semi-infinite differential induced by $d$. The
construction of these modules is as follows: Starting with the
Heisenberg and super Heisenberg algebras $h({\cal G})={\cal I}({\cal G})
\oplus {\cal E} ({\cal G}^{\prime}) \oplus Kk_{1}$ and $s({\cal
G})= \iota ({\cal G}) \oplus \epsilon ({\cal G}^{\prime}) \oplus
Kk_{2}$ with the secondary {\bf Z}-grading, we choose two subalgebras
$S({\cal G}_{\geq M} \oplus {\cal G}_{<M}^{\prime})$ and $\wedge
({\cal G}_{\geq N} \oplus {\cal G}_{<N}^{\prime})$ as the algebras of
{\it annihilation operators} with trivial representations $K \nu_{S}$,
$K \nu_{\wedge}$. Then $S^{\infty /2} {\cal G}^{\prime}$ and
$\wedge^{\infty /2} {\cal G}^{\prime}$ are defined to be the induced
representations for $h({\cal G})$ and $s({\cal G})$ respectively, and
are isomorphic (by PBW) to the algebras of {\it creation operators}
$S({\cal G}_{<M} \oplus {\cal G}_{\geq M}^{\prime})$ and $\wedge
({\cal G}_{<N} \oplus {\cal G}_{\geq N}^{\prime})$, which commute with
each other and act freely on the {\it vacuum} $\nu =\nu_{S} \otimes
\nu_{\wedge}$. ($M=N=0$ is usually referred to as the standard choice
of vacuum.
 This convention will be in effect for the rest of this
example.)

An algebra more specific than $Y^{\infty} {\cal G}$, namely
$$A^{\infty} {\cal G} = {\cal U}^{\infty} (h({\cal G}) \oplus s({\cal
G}))/(k_{1}=k_{2}=1),$$ will be home to our operators. It is well-known that
the
cocycles corresponding to the two realizations $$\underline{\pi} (x)=
\sum_{i} : {\cal I} ([x,e_{i}]) {\cal E} (e_{i}^{\prime}) :$$ and
$$\underline{\rho} (x)= \sum_{i} : \iota ([x,e_{i}]) \epsilon
(e_{i}^{\prime}) :$$ of ${\cal G}$ cancel [2], hence
$\underline{\theta} (x)=\underline{\pi} (x)+ \underline{\rho} (x)$
automatically induces a representation of ${\cal G}$ in the Weil
complex. There is a degree derivation on $A^{\infty} {\cal G}$
represented by $$\underline{\deg} = \sum_{i} i: {\cal I} (e_{i})
{\cal E} (e_{i}^{\prime}) : + \sum_{i} i: \iota (e_{i}) \epsilon
(e_{i}^{\prime}) :$$ which multiplies monomials by their (secondary)
degrees. By Lemma 2 and its symmetric version, it suffices to check
the formula only on ${\cal I}({\cal G}) \oplus \iota ({\cal G})$ !
(Normal ordering for the symmetric case does not require a factor of
$-1$.) This operator commutes with the differential and hence the
cohomology $H^{\ast} (W^{\infty /2} {\cal G},d)$ of the
complex is the direct sum of cohomologies of the eigenspaces of $\deg$ for
nonpositive integer eigenvalues.
The calculation of $H^{\ast} (W^{\infty /2}{\cal G},d)$ for arbitrary ${\cal
G}$ is in
general an extremely difficult problem, but using the techniques developed in
this paper it is not too hard to prove the following results:
\vskip 0.8truecm
PROPOSITION. {\it For any} {\bf Z}-{\it graded tame Lie algebra}
${\cal G}$ {\it we
have} $$((W^{\infty /2} {\cal G})_{\deg =0},d) \cong (W{\cal
G}_{0},d)$$
{\it and}
$$H^{\ast} ((W^{\infty /2} {\cal G})_{\deg =0},d) \cong
H^{\ast}(W{\cal G}_{0},d).$$
\vskip 0.8truecm
COROLLARY. {\it Let} ${\cal G}$ {\it be the Witt algebra} $\oplus_{n}
KL_{n}$ {\it where} $[L_{m},L_{n}]=(m-n)L_{m+n}$. {\it Then} $$ H^{
\ast} (W^{\infty /2} {\cal G},d) \cong W(KL_{0}).$$
\vskip 0.8truecm
{\it Remark.}
We know that $H^{\ast} (W \ell ,d)$ is exactly $$(S
\ell^{\prime})^{\ell} \otimes (\wedge \ell^{\prime})^{\ell}$$ for a
finite dimensional reductive Lie algebra $\ell$, so the Proposition
gives us a start for cohomological computations in a lot of
interesting cases. The Corollary follows immediately from the {\it Cartan
identity} $$\underline{d} \iota (L_{0}) + \iota (L_{0})\underline{d}=
\underline{\theta}(L_{0})$$ which reduces the cohomology to the degree zero
part since
$- \theta (L_{0})$ is the degree operator. If we change the vacuum,
the cohomology changes dramatically as shown in [2].
In case of Lie algebras that do
not naturally possess an element that acts by degree, the remaining
part of the cohomology may have a very rich structure, even for the
standard vacuum.

Results concerning the $d$-cohomology of
the Weil complex of the loop algebra of $sl(2,K)$
and more will be presented in the
author's Ph.D. Dissertation (Yale 1993).

\vskip 1.5cm
\frenchspacing
REFERENCES

\item{1.} B. FEIGIN, The semi-infinite cohomology of Kac-Moody and
Virasoro Lie algebras, {\it Russ. Math. Surv.} {\bf 39} (1984), 155-156.

\item{2.} B. FEIGIN AND E. FRENKEL, Semi-infinite Weil complex and the
Virasoro algebra, {\it Comm. Math. Phys.} {\bf 137} (1991), 617-639.
Erratum: Preprint, February 1992.

\item{3.} I. FRENKEL, H. GARLAND, AND G. ZUCKERMAN, Semi-infinite
cohomology and string theory, {\it Proc. Natl. Acad. Sci. USA}
{\bf 83} (1986), 8842-8846.

\item{4.} I. FRENKEL AND Y. ZHU, Vertex operator algebras associated
to representations of affine and Virasoro algebras, {\it Duke Math.
J.} (1992), {\it to appear}.

\item{5.} M. B. GREEN, J. H. SCHWARZ, AND
E. WITTEN, ``Superstring
Theory'', Vol. I, Cambridge University Press, Cambridge, 1987.

\item{6.} V. KAC, Laplace operators of infinite-dimensional Lie
algebras and theta functions, {\it Proc. Natl. Acad. Sci. USA}
{\bf 81} (1984), 645-647.

\item{7.} V. KAC, ``Infinite Dimensional Lie Algebras'', Birkh\"{a}user,
Boston, 1984.

\item{8.} B. H. LIAN AND G. ZUCKERMAN, BRST cohomology and highest
weight vectors, {\it Comm. Math. Phys.} {\bf 135} (1991), 547-580.

\item{9.} D. P. ZHELOBENKO, S-Algebras and Verma modules over
reductive Lie algebras, {\it Soviet Math. Dokl.} {\bf 28} (1983), 696-700.

\item{10.} G. ZUCKERMAN, ``Representations of Infinite Dimensional Lie
Algebras'', Lecture notes (Yale University), Fall 1991.

\bye